\documentclass[twocolumn,prb,showpacs,multicol,amsmath,amssymb]{revtex4}
\usepackage[dvips]{graphicx}
\usepackage{graphicx}
\usepackage{dcolumn}
\usepackage{bm}
\usepackage{graphics}
\usepackage{epsfig,color}
\usepackage[normalem]{ulem} 
\usepackage{soul}

\newcommand{\be}{\begin{equation}}
\newcommand{\ee}{\end{equation}}
\newcommand{\bea}{\begin{eqnarray}}
\newcommand{\eea}{\end{eqnarray}}

\begin{document}
\title{String orders in the Luttinger liquid phase of one-dimensional spin-1/2 systems}
\author{Hadi Cheraghi$^{1}$}
\author{Majid Jafar Tafreshi$^{1}$}
\author{Saeed Mahdavifar$^{2}$}
\email[]{smahdavifar@gmail.com}
\affiliation{$^{1}$ Department of Physics, Semnan University,  35195-363, Semnan,
Iran}
\affiliation{$^{2}$ Department of Physics, University of Guilan, 41335-1914, Rasht, Iran}
\date{\today}

\begin{abstract}
Luttinger liquid (LL) phase refers to a quantum phase which emerges in the ground state phase diagram of quite often low-dimensional quantum magnets as spin-1/2 XX, XYY and frustrated chains. It is believed that the quasi long-range order exists between particles forming the system in the LL phase. Here, at the first step we concentrate on the study of correlated spin particles in the one-dimensional (1D) spin-1/2 XX model which is exactly solvable. We show that the spin-1/2 particles form string orders with an even number of spins in the LL phase of the 1D spin-1/2 XX model. As soon as the transverse magnetic field is applied to the system, string orders with an odd number of spins induce in the LL phase. All ordered strings of spin-1/2 particles will be destroyed at the quantum critical transverse field, $h_c$. No strings exist in the saturated ferromagnetic phase. At the second step we focus on the LL phase in the ground state phase diagram of the 1D spin-1/2 XYY and frustrated ferromagnetic models. We show that the even-string orders exist in the LL phase of the 1D spin-1/2 XYY model but in the LL phase of the 1D spin-1/2 frustrated ferromagnetic model we found all kind of strings.  In addition, the existence of a clear relation between the long-distance entanglement and string orders in the LL phase is shown. Also, the effect of the thermal fluctuations on the behavior of the string orders is studied. 

\end{abstract}
\pacs{03.67.Bg; 03.67.Hk; 75.10.Pq}
\maketitle

\section{Introduction}\label{sec1}

Correlated quantum state of a matter is known as the
quantum phase. Each quantum phase is characterized by a proper order parameter considered as a measure of the degree of order in a quantum phase. Mostly, the choice of an order parameter is obvious, but in some cases, finding an appropriate order parameter is complicated.  It should be noted that in experiment, the quantum phase emerges by cooling materials to sufficiently low temperatures.

Matter has quantum fluctuations even at zero temperature. Therefore it can still support quantum phase transitions, with symmetry breaking and  order parameters  playing a key role. After discovery of the fractional quantum Hall systems\cite{Tsui82, Laughlin83}, it was found that there are some systems in  nature where these systems contain many different quantum phases at zero temperature with the same symmetry. In principle, the symmetry protected topological phases cannot be described by broken symmetries and the associated order parameter.

Among quantum phases, the study of the gapless Luttinger liquid (LL) phase has attracted much interest in recent years. We have to mention that the LL is also the paradigm for the description of interacting one-dimensional quantum systems\cite{Voit95, Imambekov12}. In this phase, the two-point correlation functions decay as power laws and the ground state is expected to have quasi-long-range order. In addition, the energy spectrum is gapless.  Based on the recent studies, it is suggested that the Wilson ratio, $R_{W}=\frac{4}{3}(\frac{\pi k_{B}}{g \mu_{B}})^{2} \frac{\chi}{C_{V}/T}$, is a crucial parameter for characterizing the LL region\cite{Wilson75, Hewson97, Jonston00, Guan13-1, Guan13-2, Guan14, Saghafi15}.  In this relation, $\chi$ is the magnetic susceptibility and $C_{V}$ is the specific heat. It was illustrated that $R_W$ is $2$ for the spin-1/2 Kondo lattice\cite{Wilson75} and the isotropic antiferromagnetic spin-1/2 chains\cite{Jonston00}.   For a system of non interacting electrons in a metal\cite{Hewson97}, this relation is equal to $1$.  Moreover, it is also reported that a double peak structure appears in the low-temperature behavior of the specific heat\cite{Abouie08, Ruegg08}. However, a one-peak structure is also reported for the specific heat\cite{Titvinidze03}. Recently, one of the authors proposed a non-local order parameter (called F-dimer order parameter) which can distinguish between a gapless LL phase and the gapped phases in the  ground state phase diagram of the alternating AF-F Heisenberg spin-1/2 chains\cite{Abouie08}. It is worth noticing the fact that the mentioned non-local order parameter is model-dependent and one cannot apply it to all 1D quantum magnets.

\begin{figure}[t]
\centerline{\psfig{file=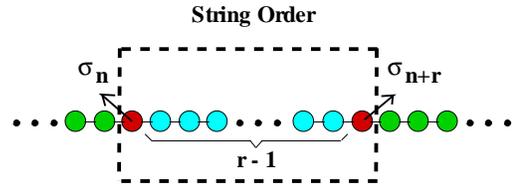,width=2.8in}}
 \caption{The schematic diagram of the string order.}\label{Fig0}
\end{figure}

In spite of all manifestations of the LL phase, the ordering of spins in this phase is not completely clear, i.e, all suggested local magnetic order parameters turn out to be zero, having no effect in determining the structure of the this phase\cite{Lake05, Balents10, Mourigal13}. The emptiness formation probability is known as a quantity which characterizes a quantum spin system in the LL phase\cite{Korepin94}. In principle, the emptiness formation probability is a probability to find a ferromagnetic string with a specified length in this phase. In this work, we continue the study on the ordering in the LL phase of the 1D quantum magnets. In particular, this phase can be found in the ground state phase diagram of the antiferromagnetic spin-1/2 chains as XX, XYY and frustrated ferromagnetic models. The zero-temperature quantum behavior of the exactly solvable spin-1/2 XX chain model shows that the ground state is in the LL phase and it undergoes a quantum phase transition by applying a transverse magnetic field\cite{Takahashi99}.

Using analytical spinless fermion approach the Hamiltonian is diagonalized. We introduce a kind of string order parameter\cite{Nijs89, Hida99} which figures out some assemblies of finite number of spins in the LL phase. The string order parameter is introduced as

\begin{eqnarray}
O(r\geq2)&=&\frac{1}{N}\sum_{n}\langle O_{n}\rangle \nonumber \\
&=&\frac{1}{N}\sum_{n}\langle[(\sigma_{n}^{x} \sigma_{n+r}^{x}+\sigma_{n}^{y} \sigma_{n+r}^{y})  \prod\limits_{m = n + 1}^{n + r - 1} \sigma_{m}^{z}]\rangle,  \nonumber \\\label{string}
\end{eqnarray}
where, $\sigma_{n}^{x, y, z}$ are Pauli operators on the $n$-th site of the chain and $\langle...\rangle$ represents expectation value on the ground state (Fig.~\ref{Fig0}). The length of each string is $r+1$. We show that in the LL phase of the spin-1/2 XX chain,  particles are correlated and they form strings with an even number of spins ($r=odd$).  Mutual interaction of the transverse magnetic field creates new correlations where particles are somehow correlated. As a result,  they form also strings with an odd number of spins. All string orders will be destroyed at the quantum critical transverse field. As a consequence, no strings exist in the saturated ferromagnetic phase. In addition to the quantum critical field, there are some fields in the LL phase of the XX chain model where the string orders disappear. We show that the mentioned string orders are related to the long-distance entanglement between the pair of particles in this model.   Additionally, we focus on the 1D spin-/12 XYY and frustrated ferromagnetic models. We show that the even-string orders exist in LL phase of the XYY model but in the LL phase of the frustrated isotropic Heisenberg model, there are all kinds of strings.

The paper is organized as follows. In the forthcoming section, we introduce the spin-1/2 XX model and present our exact analytical results for suggested string order parameter and its response to the transverse magnetic field.  We explicitly show that the long-distance entanglement can be considered as  a function of the string orders.  In Sec. III, we argue about the effect of the temperature on the string orders. In sections IV and V, we study the mentioned string orders in the LL phase of the 1D spin-1/2 XYY and  frustrated ferromagnetic models, respectively.  Finally, we conclude and summarize our results in Sec. VI.

\section{The string order}\label{sec2}

We consider  the 1D spin-1/2  XX model in the presence of a transverse magnetic field. The Hamiltonian is given as
\begin{eqnarray}
H = \frac{J}{4}\sum\limits_{n = 1}^N {(\sigma _n^x\sigma _{n + 1}^x + \sigma _n^y\sigma _{n + 1}^y) - \frac{h}{2}\sum\limits_{n = 1}^N {\sigma _n^z} }~,
\end{eqnarray}
where $J$ denotes the antiferromagnetic coupling constant and $h$ is the uniform transverse magnetic field.  This model is exactly solvable\cite{Lib61}  by using  the Jordan$-$Wigner transformation
 \begin{eqnarray}
\sigma^{\pm }_{n}&=& a_{n}^{\dag}(e^{\pm i\pi \sum_{l<n} a_{l}^{\dag}a_{l}}), \nonumber \\
\sigma^{z}_{n}&=&2 a_{n}^{\dag}a_{n}-1~,
\label{fermion operators}
\end{eqnarray}
where $\sigma_{n}^{\pm}=\frac{\sigma_{n}^{x}+i \sigma_{n}^{y} }{2}$. By performing a Fourier transformation as ${a_n} = \frac{1}{{\sqrt N }}\sum\limits_{k} {{e^{ - ikn}}} {a_k}$, the diagonalized Hamiltonian is given by
\begin{equation}
H = \sum\limits_k {\varepsilon(k)a_k^\dag } {a_k}~,
\end{equation}
where the energy spectrum is
\begin{equation}
\varepsilon(k) = J\cos (k) - h~,
\end{equation}
The ground state corresponds to the configuration in which all the states with $\varepsilon(k)<0$ are filled and $\varepsilon(k)>0$ are empty. Fermi points are given by  $\pm k_{F}=\pm \arccos(\frac{h}{J})$. One should note that the ground state of the system remains in the LL phase up to the critical transverse field $h_c=J$, where the fermionic energy band is completely filled. The spin-spin correlations in this model were studied in the classic paper by Lieb, Schultz and Mattis\cite{Lib61}. The longitudinal spin-spin correlation function is
\begin{eqnarray}
\langle \sigma_{n}^{z} \sigma_{n+r}^{z}\rangle \sim r^{-2},
\end{eqnarray}
and the transverse correlation function decays more slowly
\begin{eqnarray}
\langle \sigma_{n}^{x} \sigma_{n+r}^{x}\rangle = \langle \sigma_{n}^{y} \sigma_{n+r}^{y}\rangle \sim r^{-1/2},
\end{eqnarray}
Here, we first focus on the string order parameter (Eq.~(\ref{string})).  Using the Jordan-Wigner transformation, the fermionic form of the string order is obtained as
\begin{eqnarray} \label{eq8-1}
\frac{1}{N}\sum\limits_n {{O_n}}  = \frac{{4{{( - 1)}^{r - 1}}}}{N}\sum\limits_k {\cos (kr)a_k^\dag {a_k}}~.
\end{eqnarray}

\begin{figure}[t]
\centerline{\psfig{file=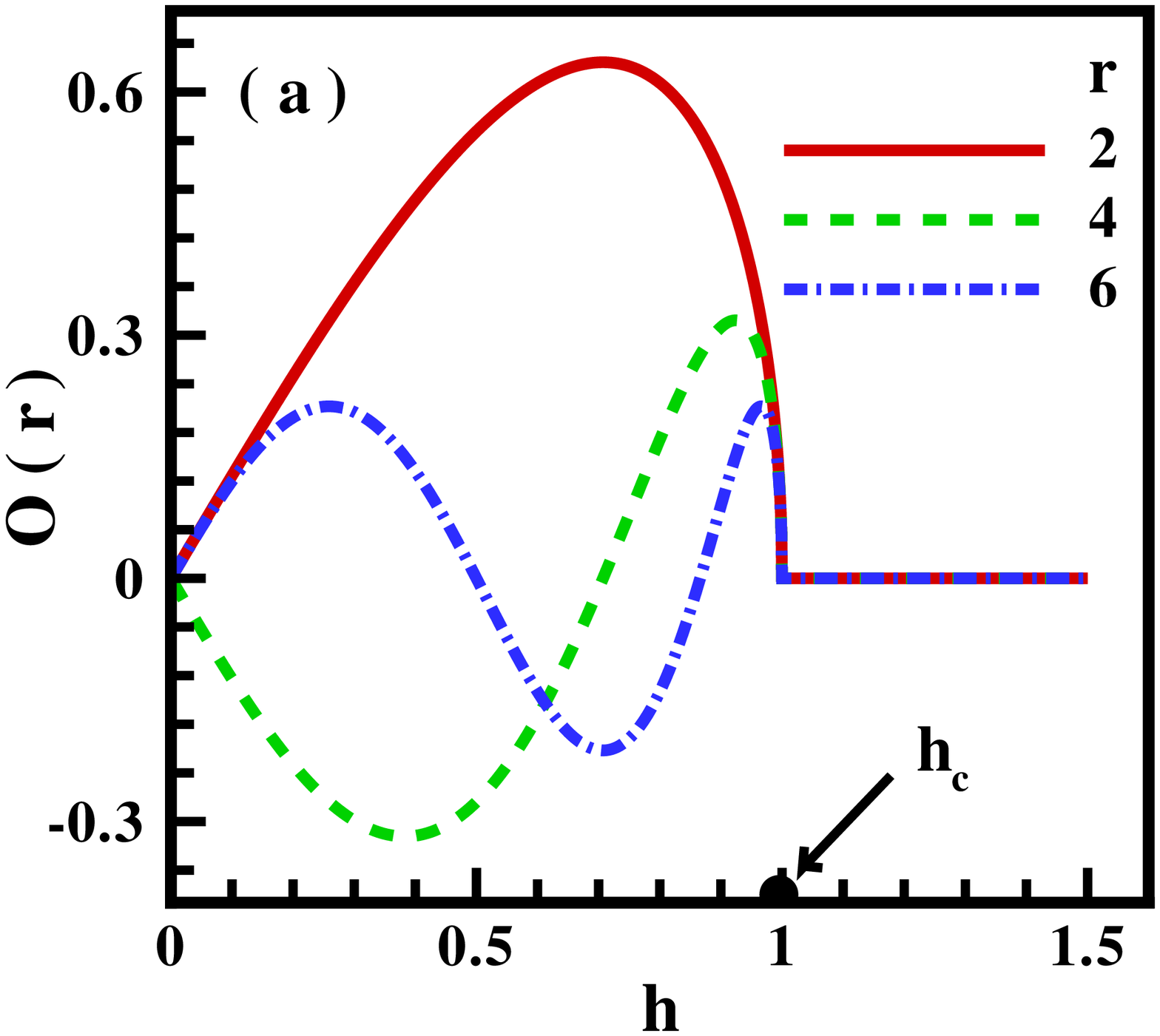,width=1.8in} \psfig{file=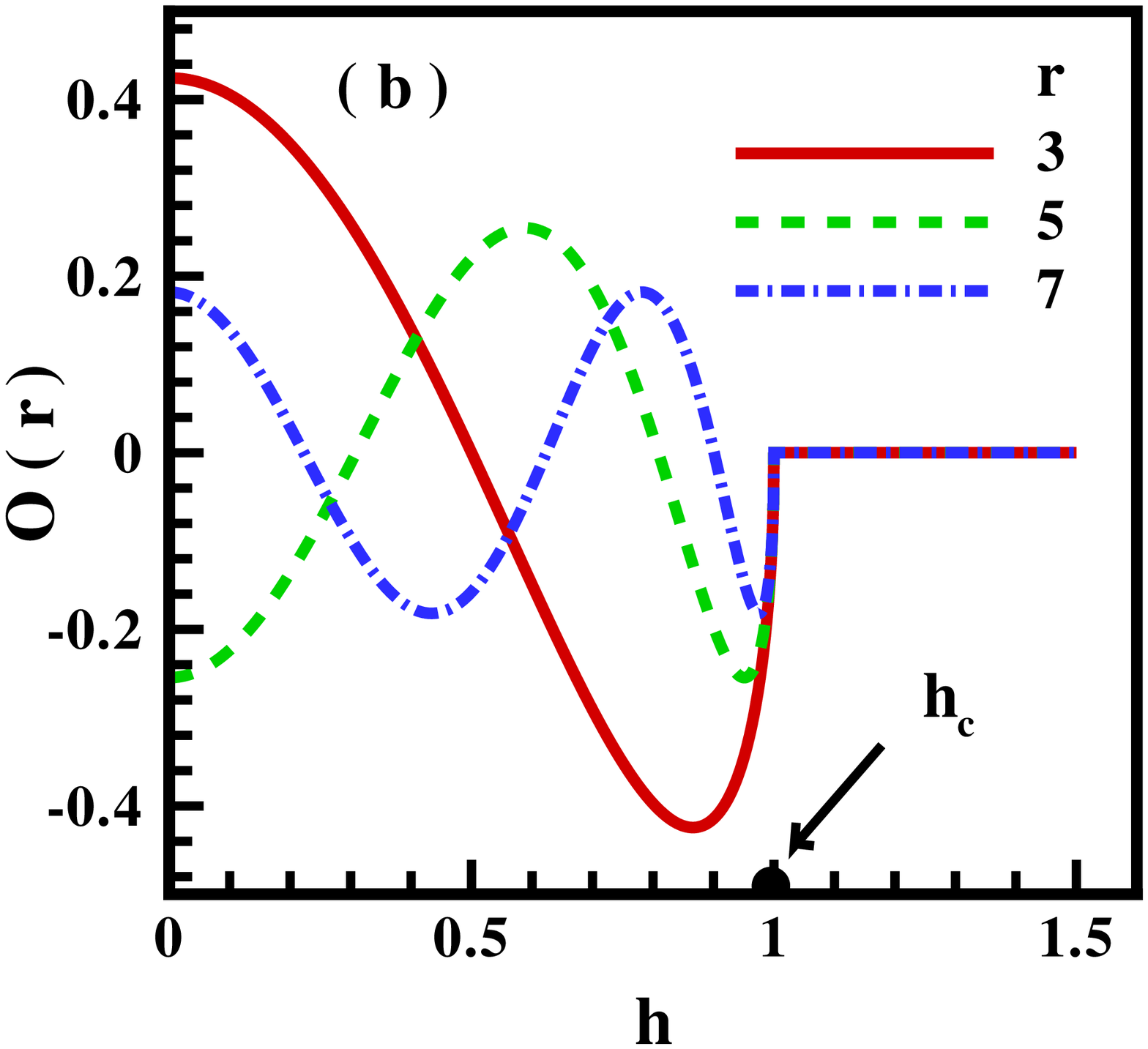,width=1.8in}}
 \caption{ (a) Odd-string orders and (b) Even-string orders versus the transverse magnetic field. Two regions are clearly separated. In the saturated FM region, $h>h_c$, no strings exist while explicitly is seen in the LL region, $h<h_c$.  }\label{Fig1}
\end{figure}

Then, taking the average on the ground state of the system the string order parameter at zero temperature is found as
\begin{eqnarray}
O(r)=\frac{4 (-1)^{r}}{\pi r} \sin(k_{F} r),
\end{eqnarray}
In the absence of the transverse magnetic field, $h=0$, the Fermi points are $\pm k_{F}=\pm \frac{\pi}{2}$. Therefore, the string order parameter is zero for   $r=even$ and $O(r)=\frac{4 (-1)^{\frac{3r-1}{2}}}{\pi r}$ for  $r=odd$. This shows that the string orders with an even number of spins exist in the LL phase of the spin-1/2 XX chain. Fig.~\ref{Fig1}  shows the string order parameter as a function of the transverse magnetic field. As is seen in Fig.~\ref{Fig1} (a), the odd-string orders   are created as soon as the transverse magnetic field is applied. Although, the  odd-string  orders increase  by increasing the magnetic field, but the  even-string orders  decrease (Fig.~\ref{Fig1} (b)),  which shows that the transverse magnetic field creates destructive quantum fluctuations on the even-string orders.  In the LL region,  all string orders will be destroyed  at some special values  of the transverse magnetic field, $h_{str}(m, r)=h_c \cos(\frac{m \pi}{r})$ where $m$ is an integer and takes $m=1, 2, ..., r-1$. It is notable that the  last value of the mentioned special transverse magnetic field coincides exactly  with the quantum critical transverse magnetic field, $h_c=J$.   By more increasing, it is clear that no string order exist in the saturated ferromagnetic phase, $h>h_c$ in complete agreement with the fundamental definition of the order parameter. At the quantum critical transverse field, $h=h_c$, the system is at the quantum critical point and the string order is zero. In the theory of the phase transition, it is believed that the thermodynamic functions show power low behavior in the vicinity of the critical point. We found that the string order shows a scaling behavior as
\begin{eqnarray}
O(r)\sim (h_c-h)^{\frac{1}{2}}~~~~~~ at~(h-h_c \longrightarrow 0_-),
\end{eqnarray}
Thus, the string order parameter exhibits the standard mean-field type behavior.


Recently, it has been suggested that the quantum entanglement can be related to the string correlations\cite{Venuti05, Miller15, Marvin17}. It is often said that the quantum orders are related to the entanglement structure of a quantum state. Hence, knowing order parameters which directly detect the entanglement structure is desirable\cite{Hamma05, Kitaev06, Levin06, Li08, Pollmann10, Jiang12}.   In the following, we show that the entanglement of formation is  related to our suggested string order parameter.

Entanglement is a specific phenomenon in which the quantum states of two or more objects have to
be described with reference to each other, even though the individual objects may be spatially separated\cite{Wooters98, Bennet00, Amico08}. Here, we focus on the long-distance entanglement between two spins of the chain model quantified by the concurrence. In fact, concurrence measures the non-local quantumness of the
correlations and  it is defined as
\begin{eqnarray}
C_{n, n+r\rq{}}&=&  \max\{0,2(|Z_{n,n+r\rq{}}|-\sqrt{X_{n,n+r\rq{}}^{+}X_{n,n+r\rq{}}^{-}})\}, \nonumber \\   \label{C2}
\end{eqnarray}
where
\begin{eqnarray}
X_{n,n+r\rq{}}^{+}&=& \langle n_{n}n_{n+r\rq{}}\rangle,\nonumber \\
Z_{n,n+r\rq{}}&=& \langle a_{n}^{\dag}(1-2a_{n}^{\dag}a_{n})(1-2a_{n+1}^{\dag}a_{n+1})\nonumber \\
&\cdots&(1-2a_{n+r\rq{}-1}^{\dag}a_{n+r\rq{}-1})a_{n+r\rq{}}\rangle,\nonumber \\
X_{n,n+r\rq{}}^{-}&=& \langle 1-n_{n}-n_{n+r\rq{}}+n_{n}n_{n+r\rq{}}\rangle,\  \label{elements}
\end{eqnarray}
In these relations, $n_{n}= a_{n}^{\dag}a_{n}$ is the occupation number operator. One can easily calculate the concurrence between the second, third and fourth neighboring spin pairs as\cite{Khastehdel16, Mahdavifar17}
\begin{eqnarray}
Z_{n,n+2}&=&f_2-2 f_0 f_2+2 f_1^{2}, \nonumber\\
X_{n,n+2}^{+}&=&f_{0}^{2} - f_{2}^{2},
\label{zx2}
\end{eqnarray}
\begin{eqnarray}
Z_{n,n+3}&=&4 (f_{1}^{3}-2 f_0 f_1 f_2+f_{2}^{2} f_1+f_{0}^{2} f_3 \nonumber\\
&-&f_{1}^{2} f_3+f_1 f_2-f_0 f_3)+f_3,\nonumber\\
X_{n,n+3}^{+}&=&f_{0}^{2} - f_{3}^{2},
\label{zx3}
\end{eqnarray}
\begin{eqnarray}
Z_{n,n+4}&=&8 (f_{1}^{4}-3 f_0 f_{1}^{2} f_2+2f_{1}^{2} f_{2}^{2}+2f_{0}^{2} f_1f_3 \nonumber\\
&+&f_{0}^{2} f_{2}^{2}-f_{2}^{4}-2f_0f_1f_2f_3+2f_1f_{2}^{2}f_3-2f_{1}^{3}f_3\nonumber\\
&+&f_{1}^{2}f_{3}^{2}-f_0f_2f_{3}^{2}-f_{0}^{3}f_4+2f_0f_{1}^{2}f_4\nonumber\\
&-&2f_{1}^{2}f_2f_4+f_0f_{2}^{2}f_4)+4(3f_{1}^{2}f_2-2f_0f_{2}^{2}-4f_0f_1f_3\nonumber\\
&+&2f_1f_2f_3+3f_{0}^{2}f_4-2f_{1}^{2}f_4+f_2f_{3}^{2}-f_{2}^{2}f_4)\nonumber\\
&+&2(2f_1f_3-3f_0f_4+2f_{2}^{2})+f_4,\nonumber\\
\nonumber\\
X_{n,n+4}^{+}&=&f_{0}^{2} - f_{4}^{2},
\label{zx4}
\end{eqnarray}
where $f_{\textit{r\rq{}}}=\langle a_{n}^{\dag} a_{n+\textit{r\rq{}}} \rangle$ and is obtained as 
\begin{eqnarray}
f_{\textit{r\rq{}}=0}&=&-\frac{k_{F}}{\pi},\nonumber \\
f_{\textit{r\rq{}}=1}&=&-\frac{1}{\pi}\sin(k_{F}), \nonumber \\
f_{\textit{r\rq{}}>1}&=&\frac{1}{4 (-1)^{r\rq{}}} O(r\rq{}),
\end{eqnarray}
which explicitly shows that the long-distance entanglement is a function of our string order parameter.
\begin{figure}[t]
\centerline{\psfig{file=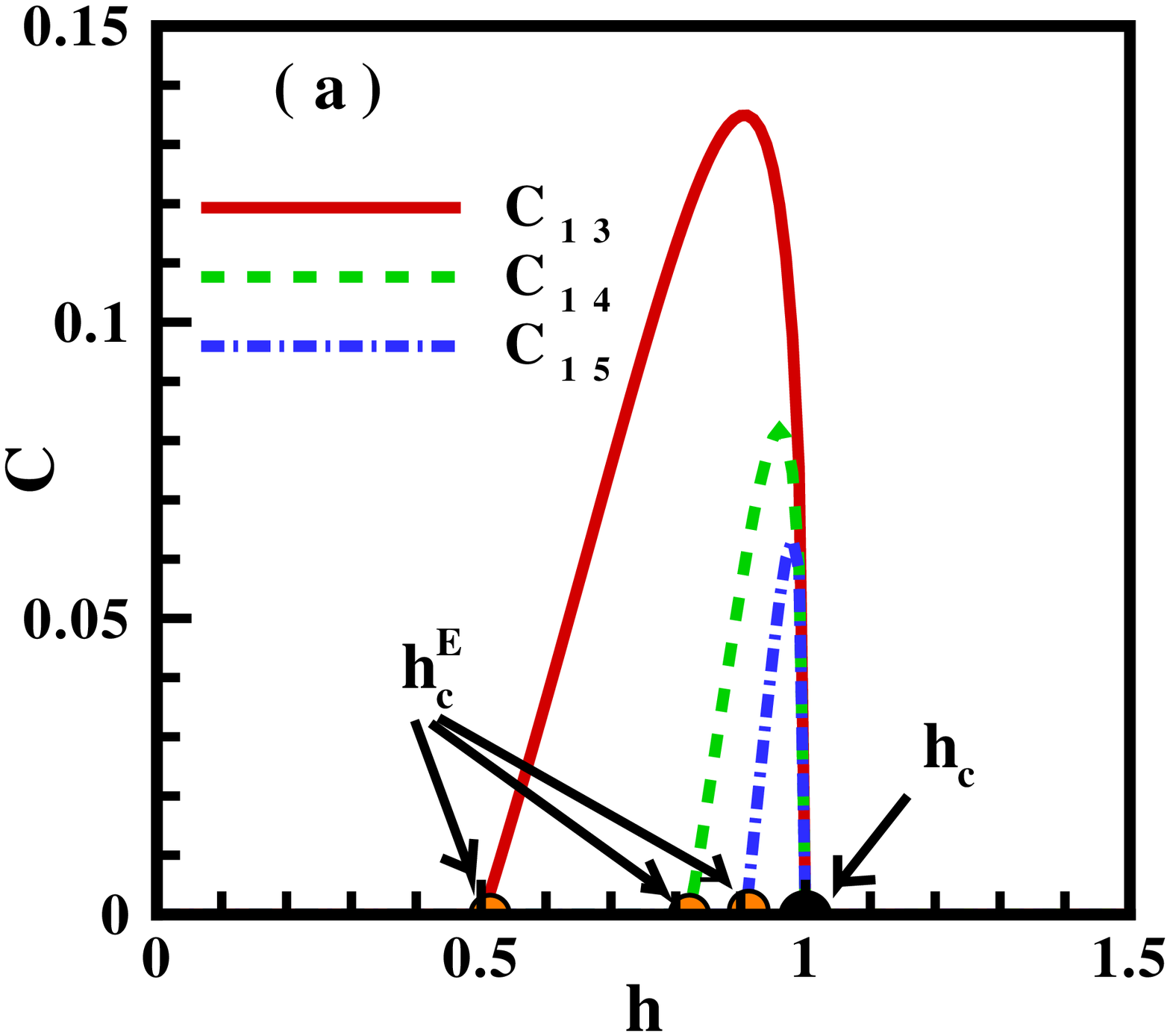,width=1.8in}\psfig{file=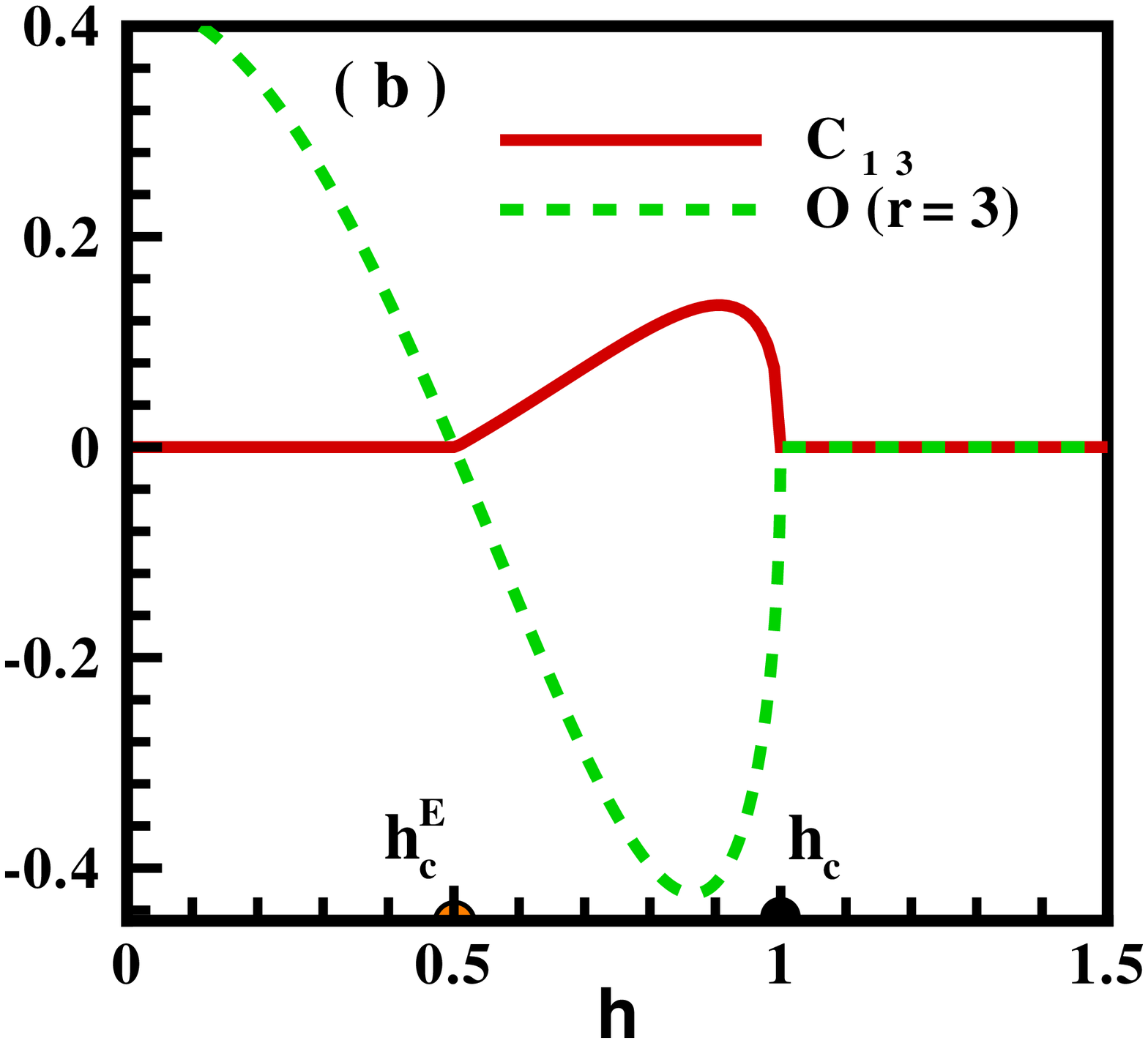,width=1.8in}}
\centerline{\psfig{file=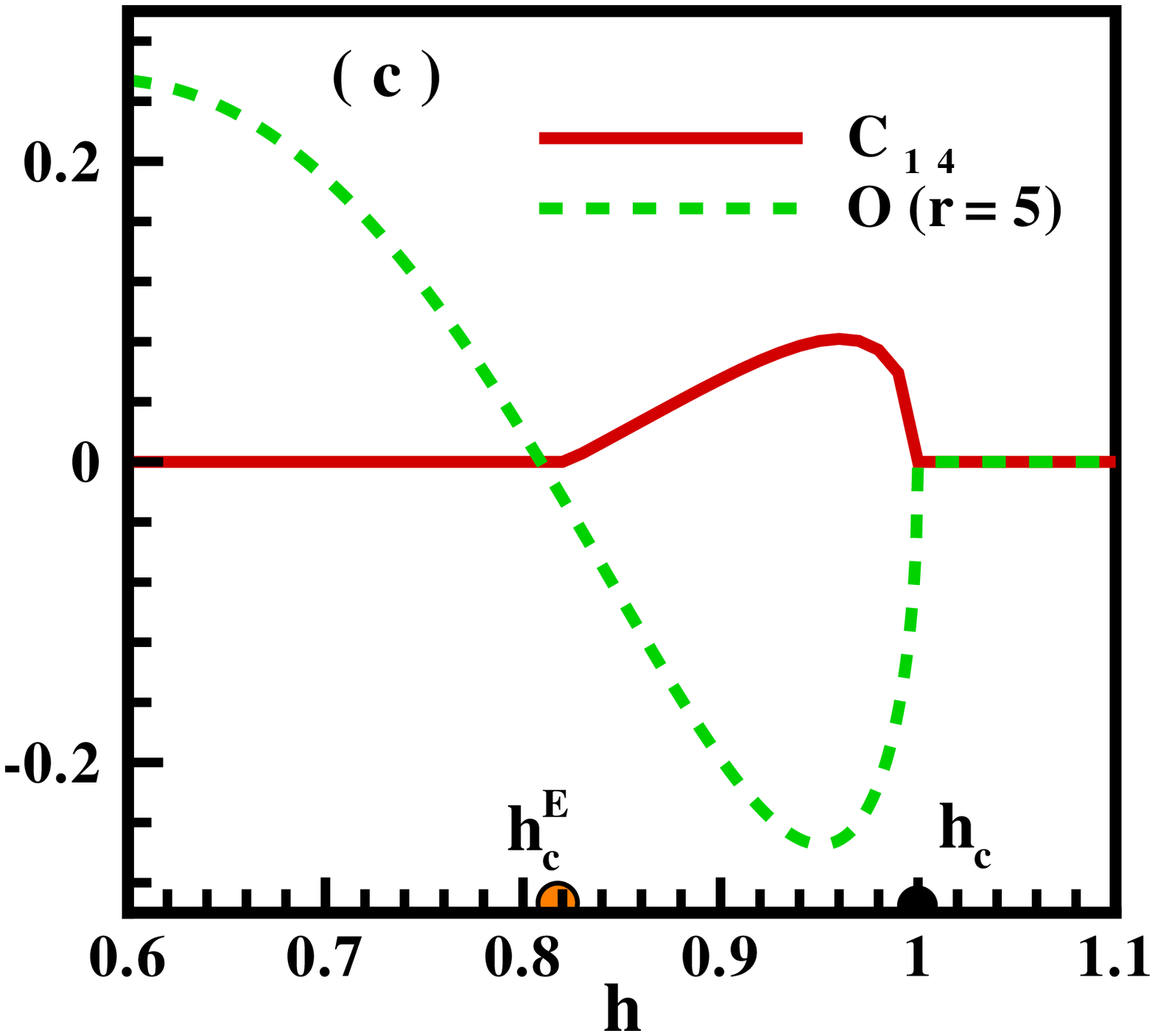,width=1.8in}\psfig{file=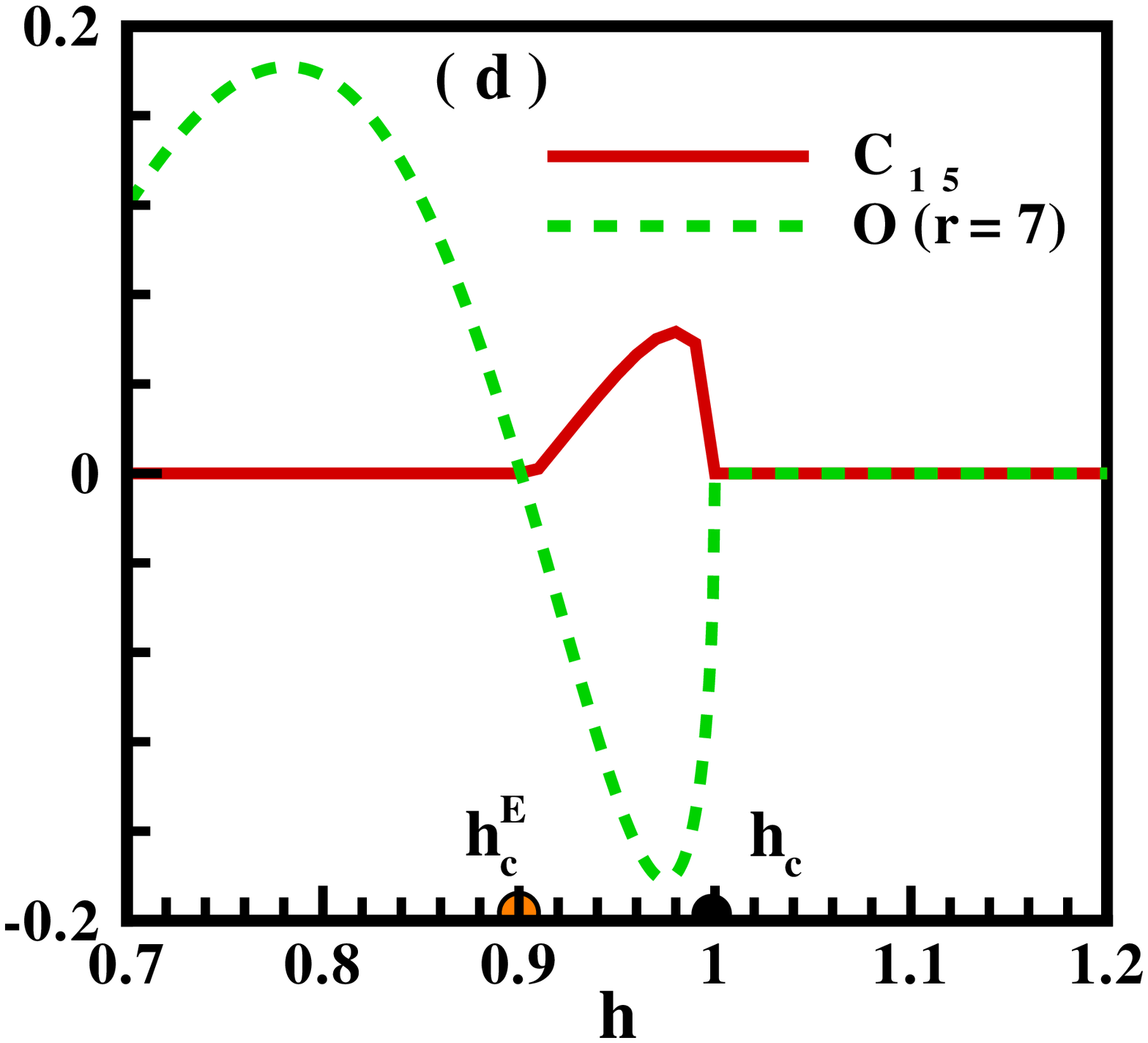,width=1.8in}}
\caption{(a) Long-distance entanglement versus the transverse magnetic field. It is clearly seen that in the LL phase, there are special regions where the particles with a distance larger than the lattice spacing are entangled. (b), (c) and (d) show the long-distance entanglement together the even-string orders.  }\label{Fig2}
\end{figure}
Results on the long-distance entanglement are presented in Fig.\ref{Fig2} (a). As is clearly seen, no long-distance entanglement exists in the LL phase of pure 1D spin-1/2 XX model. Pair spins with a distance larger than the lattice spacing remain unentangled up to a specific value of the transverse magnetic field, $h_c^{E}$, which is named  critical entangled-field and it depends on the distance between pair spins. In fact, in part of LL region which is started from  $h_c^{E}(r\rq{})$, the entanglement is created between the pair of spins with long-distance. This area will continue up to the quantum critical field, $h_c$. Any pair of spins are entangled in the saturated ferromagnetic field, $h>h_c$. In Fig.\ref{Fig2} (b), 2(c), 2(d), we have plotted the long-distance entanglement with the even-string order parameters. Here, explicitly seen that the critical entangled-field for an arbitrary distance value, $h_c^{E}(r\rq{})$, is in principle a specific value of the transverse field where the even-string order for $r=r\rq{}+1$ will be zero, which means $h_c^{E}(r\rq{})=h_{str}(1, r\rq{}+1)$. In addition, in all field-entangled regions, $h_c^{E}(r\rq{})<h<h_c$, even-string orders behave exactly as mirror images of the long-distance entanglement. Interestingly,  the maximum long-distance entanglement happens at a value of the transverse magnetic field where the even-string order parameter takes its minimum value.

\begin{figure}[t]
\centerline{\psfig{file=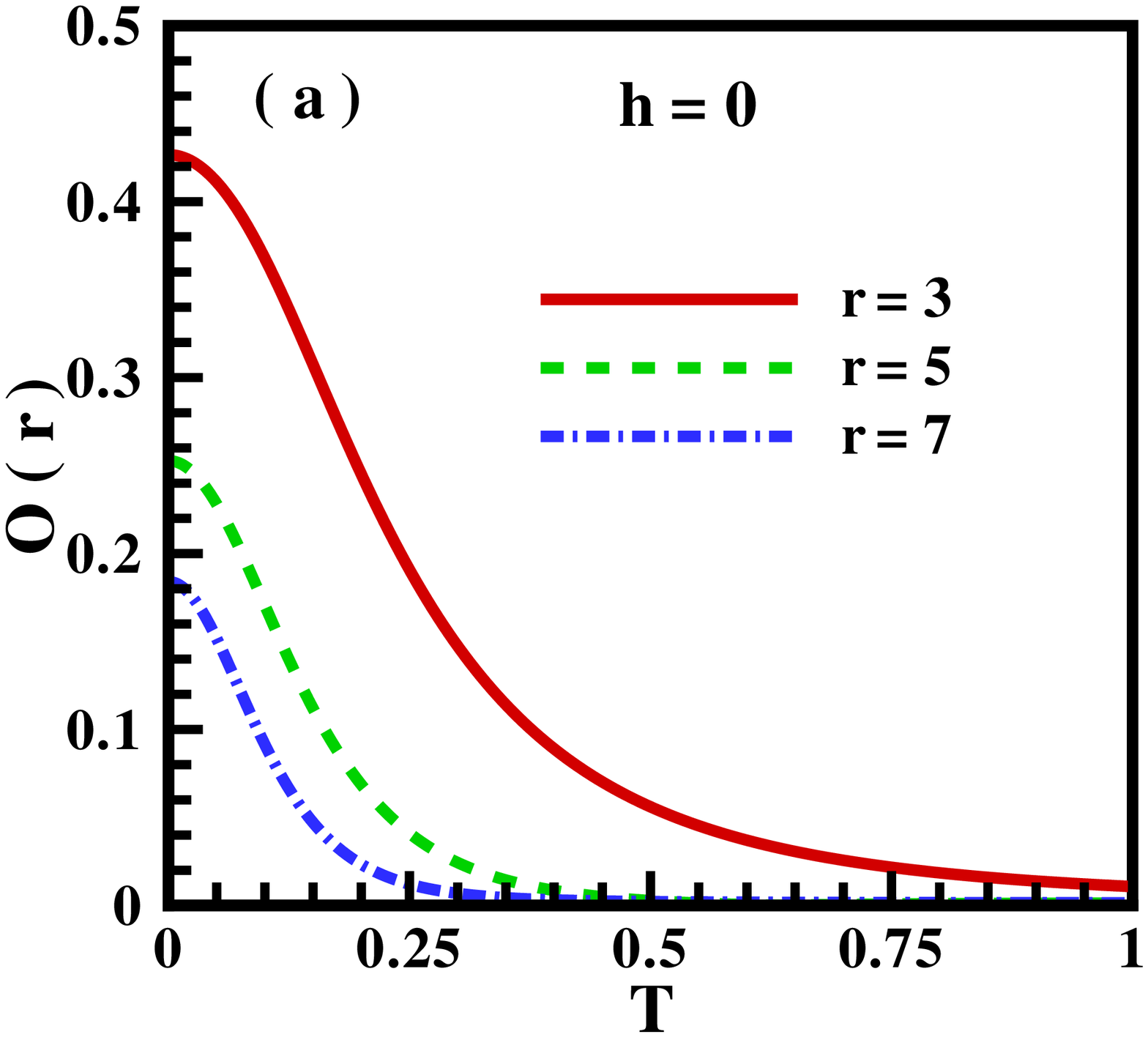,width=1.8in}\psfig{file=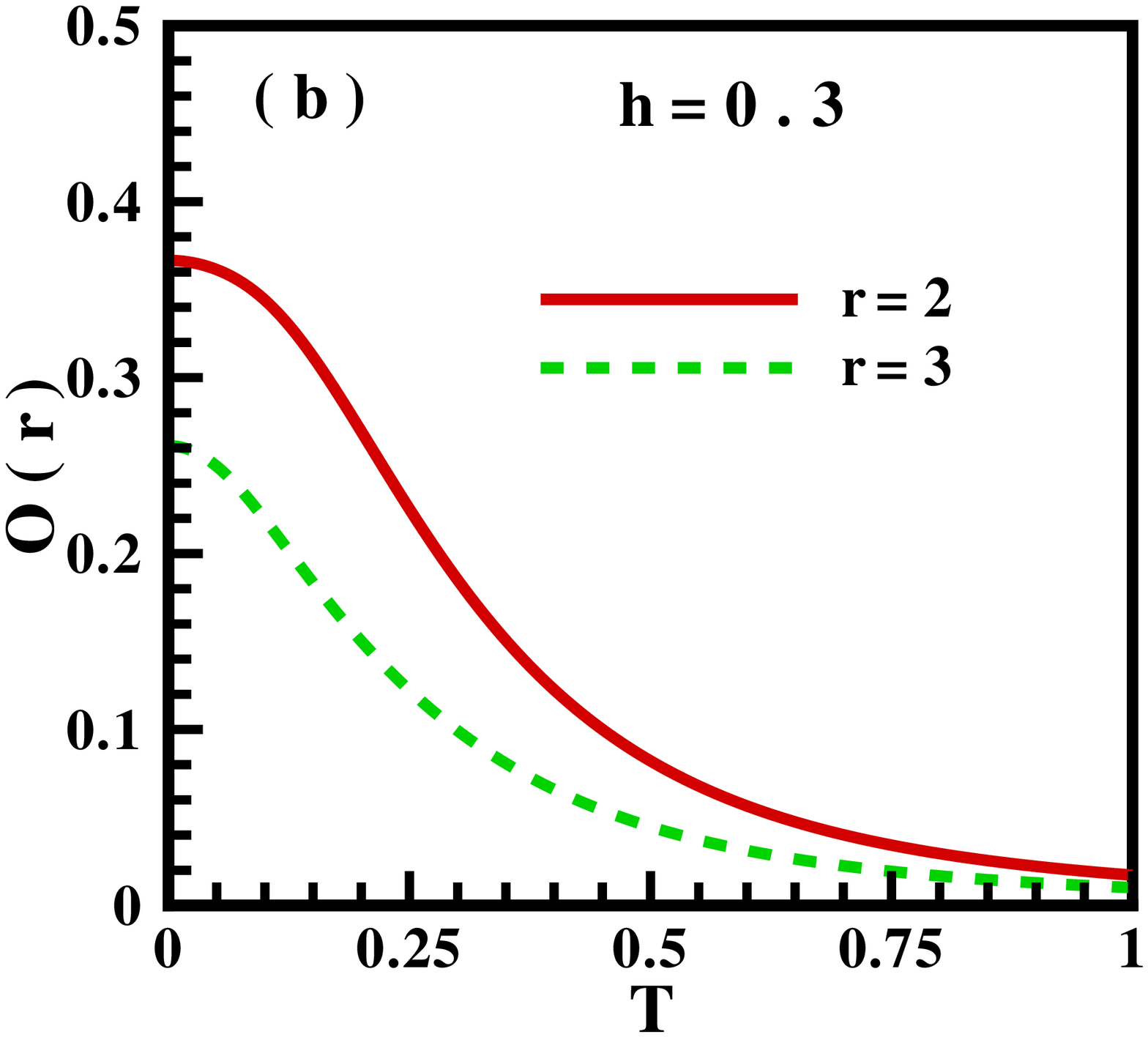,width=1.8in}}
\centerline{\psfig{file=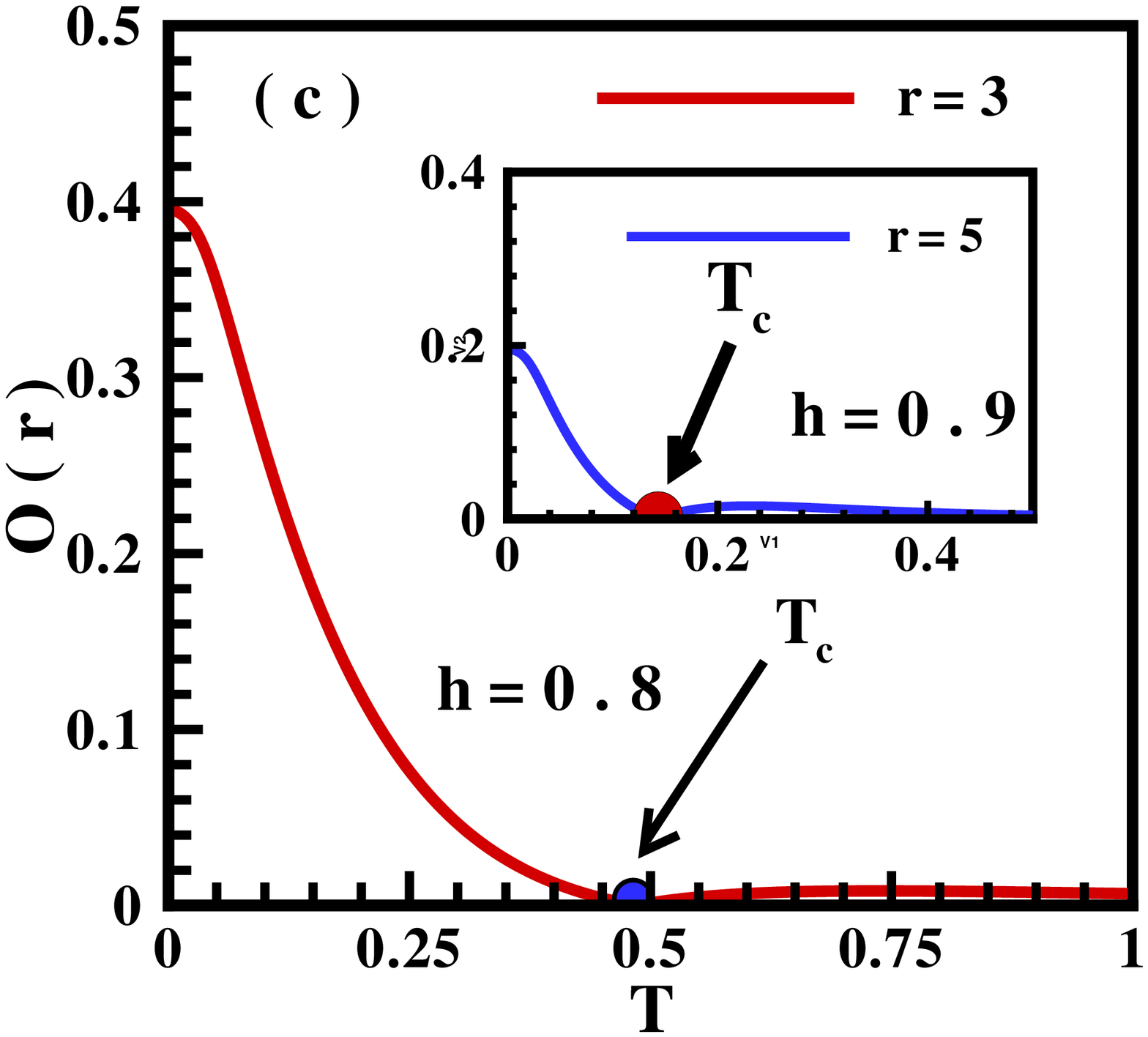,width=1.8in}\psfig{file=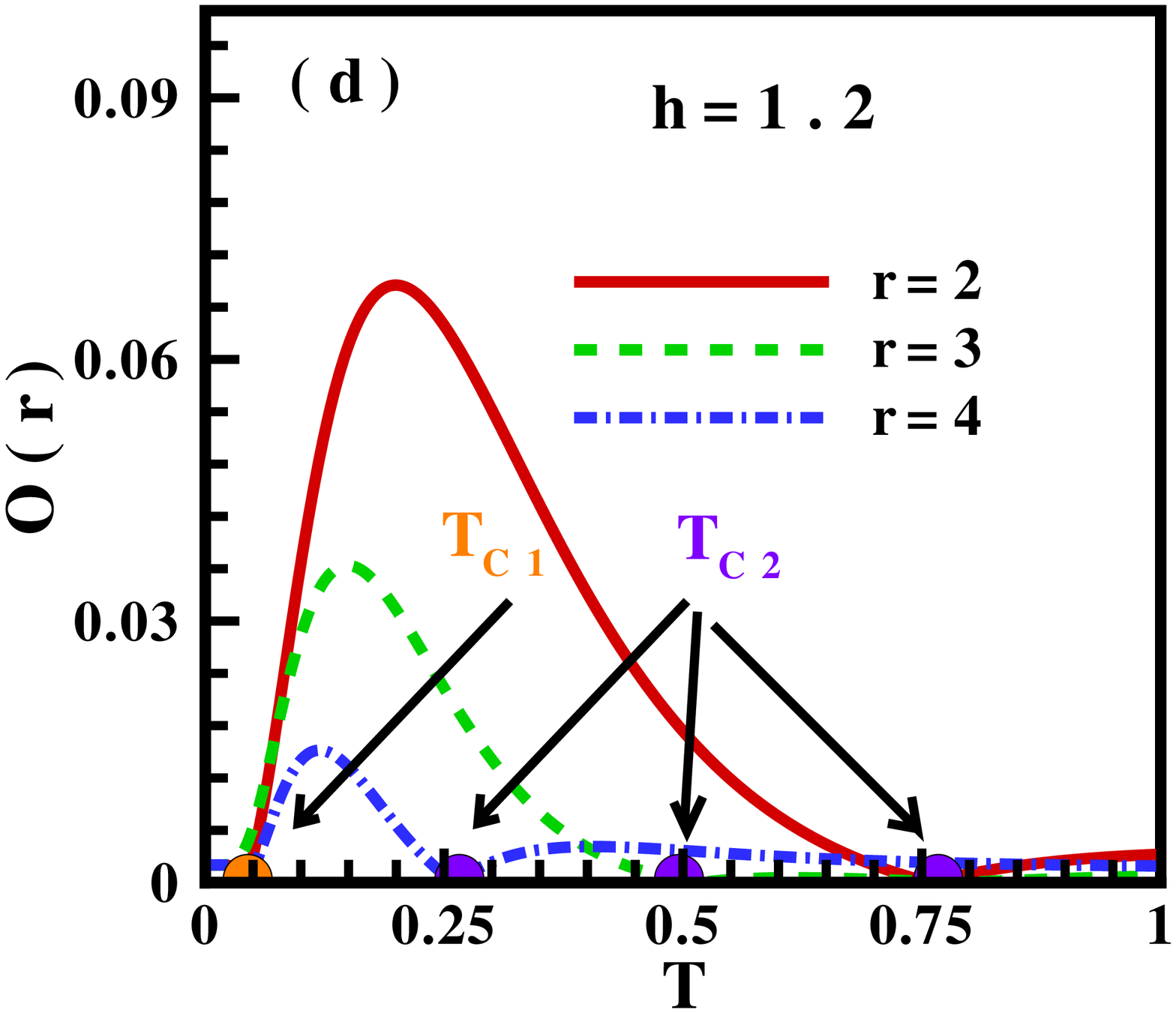,width=1.8in}}
\caption{(a) The thermal behavior of even-string orders in the absence of the transverse magnetic field. (b) The thermal behavior of the string orders in presence of the magnetic field, $h<h_{str}$. (c) In the region $h_{str}<h<h_{c}$, a critical temperature $T_c$ where the string order will be zero is clearly seen. (d) In the saturated FM  phase, $h>h_c$, no string order exist up to the first critical temperature, $T=T_{c_{1}}$. At the temperature in the interval, $T_{c_{1}}<T<T_{c_{2}}$ the string orders are created. }\label{Fig3}
\end{figure}

\section{The thermal behavior}\label{sec3}

It is known that in a system at non-zero temperature, there is the thermal fluctuations. In particular, by increasing temperature, thermal fluctuations rise and all correlations (quantum and classical) will be destroyed at sufficiently high temperature. The effect of the temperature on the quantum correlations is of particular interest and it demonstrates that the non-local correlations persist even in the thermodynamic limit\cite{Arnesen01}. Fortunately, the thermal entanglement can be experimentally verified by measurable macroscopic parameters\cite{Ghosh03, Vedral03}. Recently, the thermal long-distance entanglement in the LL phase of the 1D spin-1/2 XX model has been studied\cite{Gong09, Khastehdel16}. It is found that the long-distance entanglement remains stable up to a critical temperature.  In fact, the entanglement spreads step by step to farther neighbors with the reduction of temperature. Here, we showed that the long-distance entanglement is a function of our string order parameter. 

In the following, we focus on the thermal behavior of the string order parameter to find a better insight into the nature of the string order parameter in the LL phase.
At a finite temperature, the thermodynamic average value of  the string order parameter is obtained as $< O > = tr (\rho~O)$ where   $\rho  = \frac{{{e^{ - H/(K_B T)}}}}{Z}$ is the density matrix and the Boltzmann constant is taken as  $K_B=1$. Using the fermionic form of the diagonalized Hamiltonian, the partition function $Z = tr({e^{ - \beta H}})$  is calculated as

\begin{eqnarray}
Z ={ \prod\limits_{k}} {{\left( {1 + {e^{ - \beta \varepsilon({k})}}} \right)}}.
\end{eqnarray}

Finally the thermal string-order parameter will be  obtained as
\begin{eqnarray}
O(r)=\frac{2 (-1)^{r-1}}{\pi}\int_{-\pi}^{\pi} \frac{\cos(k r)}{1+e^{\beta \varepsilon(k)}} dk~.
\end{eqnarray}
There is something to be mentioned that the Fermi distribution
function is $\frac{1}{1+e^{\beta \varepsilon(k)}}$. Fig.~\ref{Fig3} shows the temperature dependence of the string order parameter. Since there are two phases at zero temperature, we have studied the low-temperature behavior of the string order in the regions $h<h_c$ and $h>h_c$. In the absence of the transverse magnetic field, $h=0$, as soon as the temperature increases from zero, the even-string order decreases due to the destructive effect of the thermal fluctuations (Fig.~\ref{Fig3} (a)) and
decays asymptotically at sufficiently high temperatures.  The graph in Fig.~\ref{Fig3} (b) provides information about the effect of the transverse field on the thermal behavior of the string order in the region $h<h_{str}(r)$. As it is clearly seen, all string orders decay in presence of the temperature and show the same behavior with respect to the absence of the field. More interesting results are observed in the region $h_{str}<h<h_c$ as depicted in Fig.~\ref{Fig3} (c).  It can be clearly seen, for values of the transverse magnetic field less than the quantum critical point, the string order decreases with increasing the temperature and vanishes at  critical temperature $T_c$. Physically, the number of excited states involved depends on temperature, where more states are added as the temperature is raised.  Mixing of excited states with the ground state of the system act as a destructive noise that reduces the amount of the string order in the system. When the temperature reaches the certain critical value, which varies based on the system characteristics and parameter values (as shown in the inset of Fig.~\ref{Fig3} (c)), the amount of noise created by the excited states due to thermal fluctuations is sufficient to destroy the string order of the LL phase,  only in the region $h_{str}<h<h_c$ .

On the other hand, for values of the transverse field more than the critical point, $h>h_c$, no string orders are seen at zero temperature in Fig.~\ref{Fig3} (d) in complete agreement with the saturated ferromagnetic phase. By increasing the temperature from zero, the string order does not exist up to the first critical temperature, $T_{c_{1}}$. As soon as the temperature increases from, $T_{c_{1}}$, all string orders regain and take a maximum value and then decrease and reach to zero at the second critical temperature, $T_{c_{2}}$. Since sufficiently large thermal fluctuations, the existence of the second critical temperature for destroying the string orders is completely natural. It is seen that the amount of the first critical temperature is independent of the string's length, but $T_{c_{2}}$ decreases by increasing the length of the strings. Revival of the string order in the saturated ferromagnetic region is in similar to the observed behavior of the long-distance entanglement\cite{Khastehdel16, Gong09, Mehran14} which confirms the relation between the string order and the entanglement.


\section{The string order in the spin-1/2 XYY chain}\label{sec5}

In this section we consider the 1D spin-1/2 XYY chain. The Hamiltonian  is given by

\begin{eqnarray}\label{eq17}
\begin{array}{l}
H =\frac{J}{4} \sum\limits_{n = 1}^N {\left[ {\Delta \sigma_n^x \sigma_{n + 1}^x + \left( {\sigma_n^y \sigma_{n + 1}^y + \sigma_n^z \sigma_{n + 1}^z} \right)} \right]},  
 \end{array}
\end{eqnarray}
where $J > 0$ denotes the antiferromagnetic exchange coupling  and $\Delta $ is the anisotropy parameter. The ground state phase diagram is known exactly \cite{Takahashi99}. At zero temperature  the model is integrable. In the region $ - 1 < \Delta  \le 1$, the ground state is in the gapless LL phase. The gapped Neel ordered phase exists in the $\Delta>1$ and all spins are aligned along anisotropy axis in the region $\Delta<-1$ which is called saturated ferromagnetic phase. 

Applying the Jordan-Wigner transformation (Eq.(\ref{fermion operators})), the Hamiltonian of the XYY model converts from spin operators into the spinless fermionic operators as

\begin{equation}
\begin{array}{l}
H = J(\frac{{\Delta - 1}}{4})\sum\limits_n {\left( {a_n^\dag a_{n + 1}^\dag  + h.c.} \right)} \\
\quad +J (\frac{{\Delta + 1}}{4})\sum\limits_n {\left( {a_n^\dag {a_{n + 1}} + h.c.} \right)} \\
\quad + J\sum\limits_n { {a_{n + 1}^\dag} {a_{n + 1}} {a_n^\dag} {a_n } } \\
\quad - J\sum\limits_n {a_n^\dag {a_n}} + constant.
 \end{array} 
\end{equation}

Using  Wick's theorem, the fermion interaction term is decomposed by some mean field order parameters which
are related to the two-point correlation functions as

\begin{equation}
\begin{array}{l}
{\gamma _1} =\langle {a_n^\dag {a_n}} \rangle, \\
{\gamma _2} = \langle {a_n^\dag {a_{n + 1}}} \rangle, \\
{\gamma _3} = \langle {a_n^\dag {a_{n + 1}^\dag}} \rangle.
\end{array}
\end{equation}
Finally, performing  a Fourier transformation and Bogoliubov transformation as

\begin{equation}\label{eq21-1}
{a_k} = \cos ({\theta _k}){\beta _k} + i\sin ({\theta _k})\beta _{ - k}^{\dag}~,
\end{equation}
the Hamiltonian will be diagonalized as
\begin{equation}
H = \sum\limits_k {{\varepsilon' (k)}\left( {\beta _k^\dag {\beta _k} - \frac{1}{2}} \right)},
\end{equation}
where the energy spectrum is
\begin{equation}
\begin{array}{l}
{\varepsilon' (k)} = \sqrt {a{{(k)}^2} + b{{(k)}^2}}~, \\
a(k) = J\left( {\frac{{\Delta  + 1}}{2} - 2{\gamma_2}} \right)\cos (k) + J(2{\gamma_1} - 1), \\
b(k) = J\left(  \frac{{\Delta  - 1}}{2}+{2{\gamma_3} } \right)\sin (k),
\end{array}
\end{equation}
 and $\tan (2{\theta _k}) = \frac{{ - b(k)}}{{a(k)}}$. The ground state of the system is  the vacuum of Bogoliubov $\left| \Omega  \right\rangle $ as $\beta \left| \Omega  \right\rangle = 0$. It should be noted that the mean field order parameters should be satisfied self-consistently 
\begin{equation}
\begin{array}{l}
{\gamma _{1}} = \frac{1}{2} - \frac{1}{{2N}}\sum\limits_k {\frac{{{a(k)}}}{{{\varepsilon' (k)}}}}~, \\
{\gamma _{2}} =  - \frac{1}{{2N}}\sum\limits_k {\cos (k)\frac{{{a(k)}}}{{{\varepsilon'(k)}}}}~, \\
{\gamma _{3}} = \frac{1}{{2N}}\sum\limits_k {\sin (k)\frac{{{b(k)}}}{{{\varepsilon' (k)}}}}~. 
\end{array}
\end{equation}
Ultimately, the string order parameter is obtained as

\begin{equation}
O(r) = \frac{{2{{( - 1)}^{r - 1}}}}{N}\sum\limits_k {\cos (kr)\left( {1 - \frac{{a(k)}}{{\varepsilon' (k)}}} \right)}. 
\end{equation}

\begin{figure}[t]
\centerline{\psfig{file=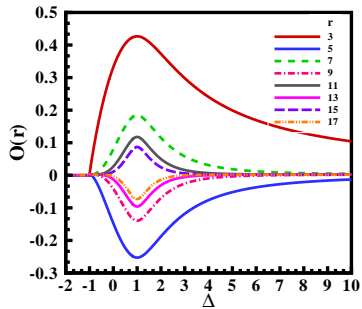,width=2.0in} }
 \caption{ (Color online.) Even-string orders as a function of the anisotropy parameter $\Delta$ in the XYY model. As is seen in the ferromagnetic phase,  $\Delta\le-1 $, no string order exists and as soon as arrive into the LL area the even-string orders are created and increased by increasing the anisotropy parameter.  }\label{Fig5}
\end{figure}

In Fig.~\ref{Fig5} we have presented our results on the string order parameter. We found that there is no odd-string order in the ground state phase diagram of the spin-1/2 XYY model. Moreover, no even-string orders exist in the saturated FM phase in the region $\Delta<-1$. As soon as the anisotropy increases from $\Delta=-1$ and in the beginning of the LL phase, the even-string order is created in the ground state of the XYY model. The even-string order parameter increases by increasing anisotropy and reaches to its maximum at the critical point $\Delta_c=1$. Recently the same behavior for the entanglement between pair of spins as a a function of the anisotropy parameter is reported\cite{Werlang10, Mahdavifar17}, which confirms existence of a relation between string order parameter and the quantum correlations.


\section{The string order in the spin-1/2 frustrated ferromagnetic chain}\label{sec6}

Another model where the existence of the LL phase in its ground state phase diagram is reported is the 1D spin-1/2  frustrated ferromagnetic chain. The Hamiltonian is given by 
\begin{eqnarray}
{ H} &=&-\frac{J_{1}}{4}  \sum_{n=1}^{N}  ({\sigma_n^x}
{\sigma}^{x}_{n+1}+{\sigma}^{y}_{n}
{\sigma}^{y}_{n+1}+ {\sigma}^{z}_{n}
{\sigma}^{z}_{n+1})\nonumber \\ 
&+&
\frac{J_{2}}{4} \sum_{n=1}^{N} ({\sigma}^{x}_{n}
{\sigma}^{x}_{n+2}+{\sigma}^{y}_{n}
{\sigma}^{y}_{n+2}+{\sigma}^{z}_{n}
{\sigma}^{z}_{n+2}),\nonumber \\  
\end{eqnarray}
where ${J}_{1}<0$ (${J}_{2}>0$) denote the nearest-neighbor and next-nearest-neighbor exchange interactions\cite{Chubukov91, Cabra00, Meisner06, Mahdavifar08, Furukawa10}. It is known that the ground state has the ferromagnetic ordering in the region $\frac{{{J_2}}}{{{J_1}}}<0.25$. In the region $\frac{{{J_2}}}{{{J_1}}}>0.25$  the ground state is in the LL phase. In the following we try to investigate the string order parameter in the ground state phase diagram of this model.

\begin{figure}[t]
\centerline{\psfig{file=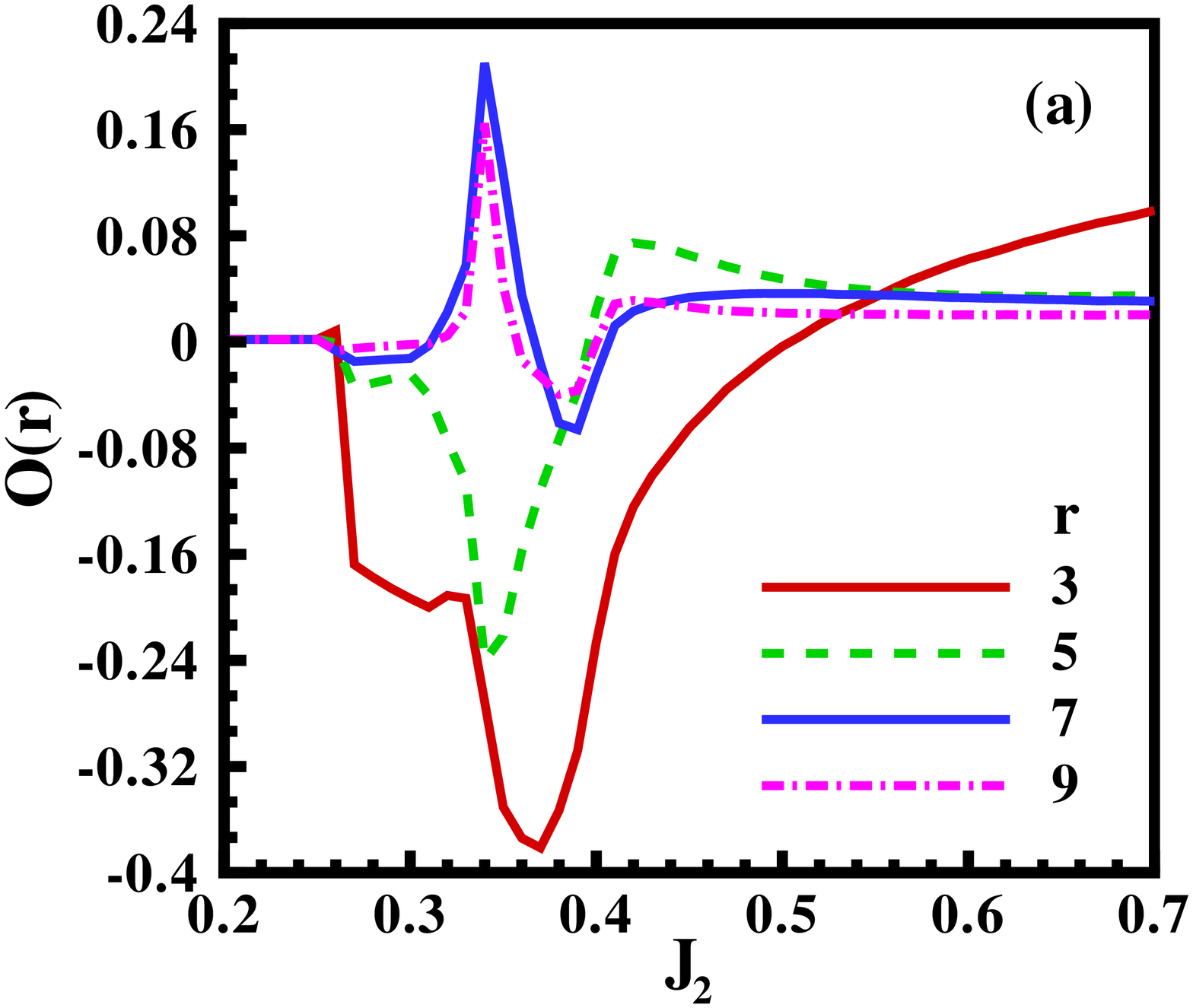,width=1.8in}\psfig{file=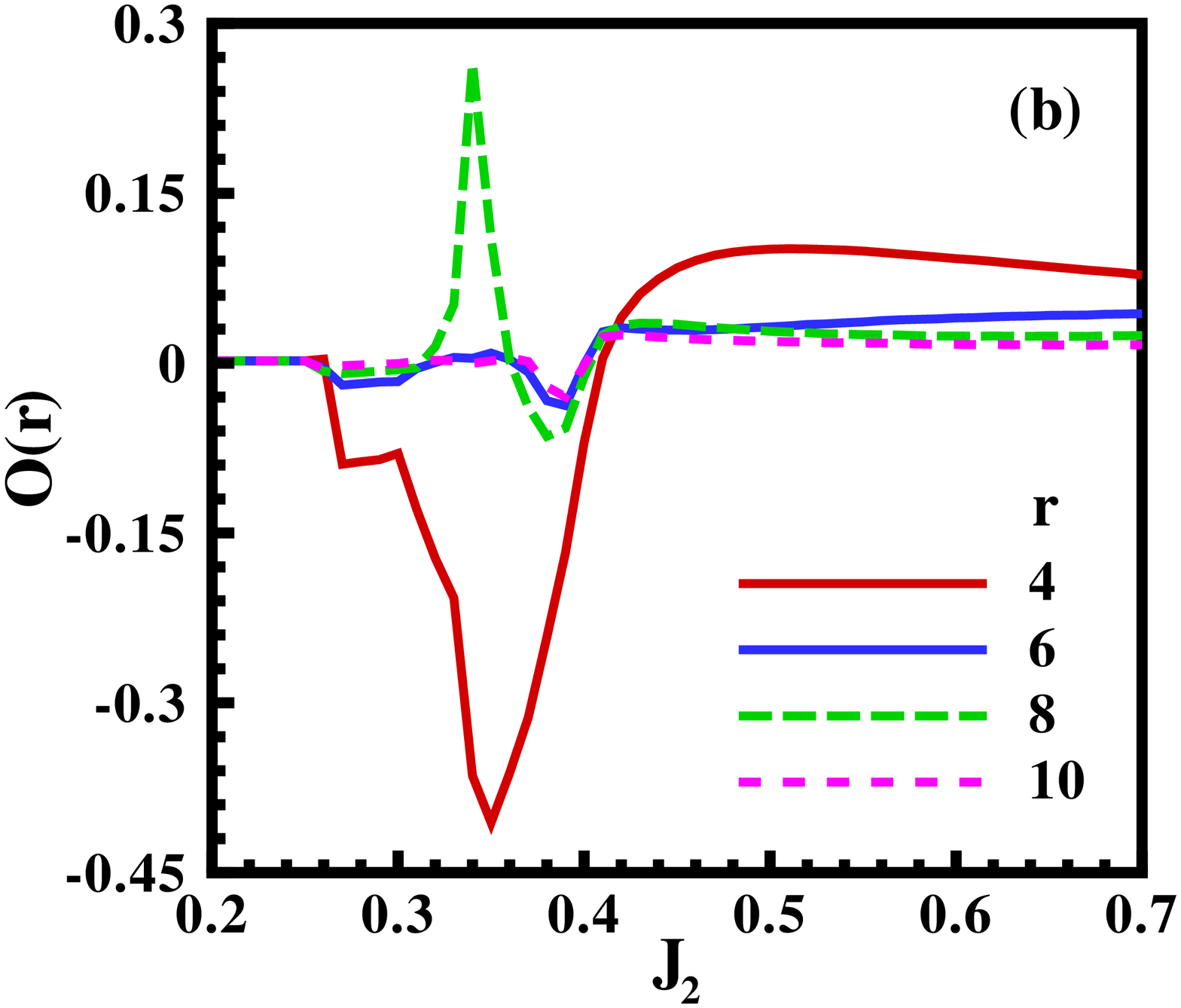,width=1.8in}}
\caption{(Color online.) String order versus next-nearest-neighbor exchange interaction, $J_2$ in the case of $J_1=1.0$: (a) Even-string orders and  (b) Odd-string orders.}\label{Fig6}
\end{figure}

First, using  the  Jordan-Wigner transformation, the fermionic Hamiltonian of frustrated ferromagnetic chain is obtained as 
\begin{eqnarray}
H&=&\sum_{n} [\frac{-J_{1}}{2}  (a^{\dag}_{n} a_{n+1}+h. c.)+\frac{J_{2}}{2}  (a^{\dag}_{n} a_{n+2}+h. c.)\nonumber \\
&+&(J_1-J_2) a^{\dag}_{n} a_{n}] \nonumber \\
&+& \sum_{n} [a^{\dag}_{n} a_{n} (J_2 a^{\dag}_{n+2} a_{n+2}-J_1 a^{\dag}_{n+1} a_{n+1})\nonumber \\
&-& J_2 (a^{\dag}_{n} a^{\dag}_{n+1} a_{n+1} a_{n+2} +h. c.)].
\label{fermionic Frustrated Hamiltonian}
\end{eqnarray}
Second, using  Wick's theorem, the fermion interaction
terms are decomposed by following mean field order parameters
\begin{eqnarray}
\gamma'_{1} &=& <a_{n}^{\dag}a_{n}>,\nonumber \\ 
\gamma'_{2} &=& <a_{n}^{\dag}a_{n+1}>,\nonumber \\
\gamma'_{3} &=& <a_{n}^{\dag}a_{n+2}>.\nonumber \\
\end{eqnarray}
Third, utilizing these order parameters and performing a Fourier transformation, the diagonalized Hamiltonian is given by 
\begin{eqnarray}
 H =  \sum_{k} \varepsilon''(k) a_{k}^{\dag}a_{k}~,  
\end{eqnarray}
where the energy spectrum is 
\begin{eqnarray}
\varepsilon''(k) &=&  A+ B \cos(k) + C \cos(2 k),\nonumber \\
A &=&(J_1-J_2)(1-2 \gamma'_{1})-2 J_{2}\gamma'_{3}~,\nonumber \\
B &=& J_1(-1+2\gamma'_{2})+4 J_2\gamma'_{2}~,\nonumber \\
C &=&  J_2[1-2 (\gamma'_{1}+ \gamma'_{3})].
\end{eqnarray}
The ground state corresponds to the configuration in which all the states with $\varepsilon''(k)<0$ are filled and $\varepsilon''(k)>0$ are empty. In the  region $\frac{{{J_2}}}{{{J_1}}} \ge 0.25$, we found four Fermi points as 
\begin{equation}
\begin{array}{l}
\pm k_F^ -  =  \pm \arccos \left( {\frac{B}{4C} - \sqrt {\frac{{{B^2}}}{16C^2} - \frac{{{A-C}}}{2C}} } \right), \\
\pm k_F^ +  =  \pm \arccos \left( {\frac{B}{4C} + \sqrt {\frac{{{B^2}}}{16C^2} - \frac{{{A-C}}}{2C}} } \right).
\end{array}
\end{equation}
One should note that the following equations should also be satisfied self-consistently
\begin{equation}
\begin{array}{l}
{\gamma'_1} = 1 - \frac{{k_F^ -  - k_F^ + }}{\pi }, \\
{\gamma'_2} =  - \frac{1}{\pi }\left( {\sin (k_F^ - ) - \sin (k_F^ + )} \right),\\
{\gamma'_3} =  - \frac{1}{2\pi }\left( {\sin (2k_F^ - ) - \sin (2k_F^ + )} \right).
\end{array}
\end{equation}
Finally, the string-order in the frustrated ferromagnetic chain  will be obtained as

\begin{equation}
O(r)=\frac{4(-1)^r}{\pi r}\left({\sin ({k_F^ -}r ) - \sin ({k_F^ +}r )} \right).
\end{equation}
The string order parameter  has been depicted  in Figs.~\ref{Fig6}(a)-(b) versus the next-nearest-neighbor exchange interaction $J_2$. As seen in Fig.~\ref{Fig6}, no odd and even-string orders exist in the ferromagnetic phase region $\frac{{{J_2}}}{{{J_1}}}<0.25$. As soon as the next-nearest-neighbor exchange interaction increases from critical value $\frac{{{J_2}}}{{{J_1}}}=0.25$, both odd and even-string orders are created in the ground state of the system. The mentioned string orders are seen in a narrow region of the LL phase of the ground state phase diagram.

\section{DISCUSSION}\label{sec7}
The study of the zero-temperature behavior of low-dimensional quantum magnets has a long history.
 Most physical information is obtained from the ground state phase diagram of a low-dimensional quantum magnet. One of the challenges of the quantum phases is known as LL phase. In particular, it is proposed that the quasi-long-range order exists in the mentioned phase. In this paper, we tried to find a better picture of the ordering between particles in the this phase. For this purpose, we considered an exactly solvable spin-1/2 XX chain in a transverse magnetic field as a model where the LL phase is seen in its ground state phase diagram. To show that the results obtained for this model can be applied to the LL phase in general, we also studied the 1D spin-1/2 XYY and frustrated ferromagnetic models.

We suggested a kind of string order parameter and showed that the spin-1/2 particles can form strings with the odd or even number of spins in the LL phase of the mentioned models. On the other hand, we showed that the mentioned string order is related to the entanglement of formation. 

Additionally, we focused on the thermal behavior of the strings at low temperatures in the exactly solvable spin-1/2 XX chain model. Interesting results are found in the saturated ferromagnetic phase. Although there are no strings between spin-1/2 particles at zero temperature, increasing temperature creates a kind of correlation between the particles which  leads to the formation of strings.

\section{acknowledgments}
The authors thank  G. I. Japaridze and M. Motamedifar for valuable comments.


\vspace{0.3cm}


\end{document}